\documentclass[english]{lni}
\pdfoutput=1
\def\ie{i.\,e.,~}

\def\lqq{\lq\lq}
\def\rqq{\rq\rq}
\def\dq#1{\lqq #1\rqq}

\usepackage{graphicx}
\title{Smart Microgrids: Overview and Outlook}
\author{Anita Sobe, Wilfried Elmenreich\\
Institute of Networked and Embedded Systems,\\
 Alpen-Adria Universit\"at Klagenfurt, Austria\\
anita.sobe@aau.at, wilfried.elmenreich@aau.at}
\begin{document}
\maketitle
\begin{abstract}
The idea of changing our energy system from a hierarchical design into a set of nearly independent microgrids becomes feasible with the availability of small renewable energy generators. The smart microgrid concept comes with several challenges in research and engineering targeting load balancing, pricing, consumer integration and home automation.
In this paper we first provide an overview on these challenges and present approaches that target the problems identified. While there exist promising algorithms for the particular field, we see a missing integration which specifically targets smart microgrids. Therefore, we propose an architecture that integrates the presented approaches and defines interfaces between the identified components such as generators, storage, smart and \dq{dumb} devices.
\end{abstract}
\section{Introduction}
The trend towards distributed renewable energy production leads to new challenges. Renewable energy sources typically rely on the weather and thus lead to variable energy production which is hard to manage. However, they are an important part of future smart grids and therefore there are a lot of efforts to make these sources more efficient. Future smart grids will not only have to integrate distributed renewable energy sources, but will also have to integrate information and communication technologies (ICT) for management and control~\cite{Farhangi2010}. Currently, ICT integration is done by installing smart meters, which opens a wide area of new applications.

Future efforts target the increase of manageability and efficiency by dividing the smart grid into sub-systems~\cite{Mohn2011a}. Such sub-systems are called \emph{smart microgrids} and consist of energy consumers and producers at a small scale and are able to manage themselves. Examples for smart microgrids are households, villages, industry sites, or a university campus. A smart microgrid can either be connected to the backbone grid, to other microgrids or it can run in a so called island mode. Dynamic islanding is one of the main solutions to overcome faults and voltage sags \cite{Lasseter2011}. According to Mohn and Piasecky in \cite{Mohn2011a} smart microgrids need to be controlled on two levels, (1) analog-centric control for power stability and (2) digital-centric control for system automation. We are specifically interested in the second level, which is responsible for calculating the need for energy based on its price, reliability, and current system state.

In more detail we will give an overview on dynamic pricing, Smart Home automation in combination with Demand Response (DR) and power load balancing in island mode. These topics target improved reliability, better management of distributed resources, and higher power efficiency, but are typically isolated research efforts. We want to subsume these topics and strive to give an outlook on a holistic approach of smart microgrids.
\section{Dynamic Pricing}
\label{sec:pricing}
Currently there are only a small number of market trader companies that manage the energy market, but with the envisioned network of smart microgrids a high number of components emerge that are influencing each other. This will lead to new business models and dynamic market models. Since it is possible to integrate multiple consumers and intermediate traders at the intersections of several microgrids, agent-based solutions emerge. An agent can be represented by a meter, a control entity, market entity, aggregation entities, etc.~\cite{Duan2011}. All of these agents are capable of communicating with each other and based on that make decisions. On the market level agents can buy and sell energy and negotiate prices. In the following we discuss two of the current approaches for price agreement - market-based approaches and game theoretic considerations.
\subsection{Market-based approaches}
Market-based approaches are efficient in dealing with different kinds of decentralized trading systems \cite{Clearwater1996}, e.g., the stock market. Advantages of market-based approaches are the local view of the agents and there is no need to publish their bids and asks before clearance. We concentrate on published market based mechanisms for energy pricing, where also small producers are considered, i.e., small photovoltaic entities, or wind generators on household, but also in shared cases such as a campus or an industry site. The market mechanism consists of day-ahead pricing and real-time pricing mechanisms and in some nations this market is even open for individual consumers. 

We found several implementations of market-based approaches and strive to give an overview by categorizing them based on the following criteria.
\begin{itemize}
 \item  Producers (level of detail): shared energy producers, homes
  \item Supports island mode: yes, no
  \item Auction protocol
  \item Capacity feedback loop: yes, no
  \item Goal of the approach
\end{itemize}
One of the well known implementations of simulating pricing in the energy market is called AMES \cite{Li2008}. In this system the authors consider a combination of pricing with capacity and rely on Locational Marginal Prices (LMP). For each node in the network, either producing or consuming energy, a separate price is calculated a day ahead of the actual consumption/production based on hourly reports. A consumer can have a fixed demand that is not price-sensitive and a dynamic demand that is price-sensitive, the same applies for the producers. The system consists of an independent system operator (ISO) that collects bids and offers from each buyer and seller a day ahead. Each hour the ISO reports the commitments based on the LMPs for the next day, allowing to adjust the dynamic demands and offers of the buyers and sellers. The authors compare different ratios of price-sensitive demands, the impact of price caps and generator learning capabilities (strategic behavior). The generator learning behavior has a high impact on the price a consumer has to pay. Even with price caps, the model might lead to significantly higher prices. The work relies on shared energy producers and central entities. It is not mentioned if islanding is supported. However, if we consider small microgrids, each of it needs a central entity that collects bids and offers also considering imbalances for renewable energy generators. The communication relies on the knowledge and reliable connection of all nodes within the network. The pricing is dependent on the capacity of a generator, however, this work does not consider congestion on selected power links.

An extension to AMES can be found in \cite{Vytelingum2010}. The authors argue that in the work of Li et al \cite{Li2008} the learning mechanisms of generators lead to unfair market conditions for the consumers and reduce the efficiency of the system. In this work, the authors use Continuous Double Auction (CDA), which is successfully applied in the stock market. Sellers and buyers can place their bids continuously during a fixed trading period. If a match between buyers and sellers is found, the auction is performed immediately. An interesting point is that the mechanism leads to price adaptations in case of congestion. The proposed mechanism consists of three parts: (1) the Trading Mechanism controls agents' interactions, (2) the Security Mechanism controls the transmission line capacities, (3) the Online Balancing Mechanism ensures fair pricing for extra demand and supply. The balancing mechanism manages the extra demand and supply based on secure quantities instead of price, \ie often leads to losses for bad predictions in the day ahead market. The authors evaluate their mechanism by applying different agent behavior strategies, namely (1) zero-intelligence, (2) AA-strategy with an extension to electricity markets, \ie AA-EM. Zero-intelligence draws a random number between 0 and the limit price for offers and bids. The AA-EM strategy considers a weighted average of transaction prices within one node to calculate the equilibrium price. In a fully connected network and even in the small world network the system efficiency goes up to 95 \%. To support islanding the mechanism relies on a trader agent and network control agent in each microgrid. The authors further assume reliable connection links for communication. The mechanism considers explicitly sellers and buyers on the same node, \ie this mechanism can be applied on household level.

Ramachandran et al. describe in \cite{Ramachandran2011} a CDA based auction model that considers islanding of microgrids (extends \cite{Dimeas2005}). The goal is to maximize the consumption from the local distributed energy resources without selling energy to the global grid. If the load demand exceeds the generation rate, energy is bought from the global grid. For the optimization method the authors propose to use bio-inspired algorithms and compare different implementations based on artificial immune systems. In artificial immune systems antigens are invaders that should be attacked by matching antibodies built as an immune-response. The first method called AIS \cite{DeCastro1999} evaluates the antibodies according to their affinity (cost). A number of antibodies is called a generation and the next generation evolves by evaluating the affinity values of each antibody. Depending on the goodness of an antibody either a copy, a mutation or a new generated antibody is selected for the next generation. The input of the immune system is the price and the number of antibodies (i.e., distributed energy resources). The authors argue that this method is inefficient because of redundant searches for the optimum and propose an improvement operator inspired by particle swarm intelligence (called IPSO). Instead of cloning the best candidates the authors propose to improve their affinity instantly. Each of the candidates is seen as a particle in the swarm and as such can have a velocity and a position. The goal is to calculate the velocity and position of that particle for the next stage. The better the position in the next stage the more likely is the cloning of this candidate. This approach shows price reductions of up to 37 \% in using DERs efficiently in comparison to only using the global grid.

Although not applied to energy markets the authors of \cite{Das2001} show interesting results when comparing the performance (gains of trading) of software agents with human traders in a laboratory auction based on CDA. In the energy market the number of agents (seller/buyers) is large, which makes the system complex. In this experiment the number of agents is limited to 12 and even in this case the software agents lead to a better performance than the human counterparts. The agent behavior follows a modified zero-intelligence-plus algorithm where period lengths and persistent open orders are considered. The authors compare ZI-plus with the Gjerstad and Dickhaut (GD)~\cite{Gjerstad1998} algorithm, which is based on historical trades and calculates a belief value. The original implementation leads to a high fluctuation of bids and asks with a high variance in the prices. Both of the methods, however, would have to be changed to be applied to the energy market (i.e., considering link capacities and congestion). A central auctioneer is considered that manages all bids and asks.

In Duan and Deconinck \cite{Duan2010} traditional auction-based methods are compared according to the number of messages that have to be exchanged. The authors further compare the increase of messages to be exchanged if the number of agents increases. The main contribution is an architecture and communication protocol for smart microgrids that allow for implementing different auction based methods. The basic market procedure consists of three phases (1) initial price, (2) bid \& adjust and (3) demand function. The auction methods are English auction, First-price sealed-bid, Dutch auction and Vickrey auction \cite{Sandholm1999}. In the English auction the bidders raise their prices until none of the bidders raises anymore. The opposite mechanism is the Dutch auction, where the bidders lower their prices until one of the bidders takes the good. In the First-price sealed-bid auction each bidder provides an uninformed bid and the highest bid wins. The Vickrey auction works similar, but the price the winner has to pay is the second-highest bid. The basic assumption of all of these methods is that there is one seller and a number of bidders. In the energy market, this setting might apply for one of the microgrids, but even there several power generators might act as sellers. Further, the authors do not consider the question on how the different microgrids should interact with each other.

Since these approaches have different goals and applications we compare them according to the categories identified above. We classify the approaches by producer types, i.e. if the households are producers themselves or if the energy source is shared, by the capability of supporting island mode, by the type of the auction protocol and if the energy capacity is considered. Finally, we state the goal of the approach. In table \ref{tab:comp} we subsume the discussion of the before mentioned approaches.
\begin{table}
  \centering
  \caption{Comparison of Market-based Approaches for Microgrids, Energy source - S:shared, H:home, Island mode -  +:yes, -:no, o:depends, auction protocol, capacity feedback +:yes, -:no, and goal}\label{tab:comp}
  \begin{tabular}{l ccccl}
  \\
    \hline
    & Source  &  Island & Auction & Capacity F. & Goal\\
    \hline \\
   \cite{Li2008}& S & o &LMP & - & market efficiency\\
   \cite{Vytelingum2010} & H &  o & CDA & + & market and DER efficiency\\
   \cite{Ramachandran2011}& H &  + & CDA & + & DER efficiency\\
   \cite{Das2001} & S & - & CDA & - & market efficiency\\
   \cite{Duan2010}&S & +  & traditional & - & communication efficiency\\ \hline
  \end{tabular}
\end{table}
\subsection{Game-theoretic approaches}
Game-theoretic approaches are a promising tool for the analysis of smart grid systems. There are numerous applications, especially in the form of non-cooperative games and learning algorithms. Non-cooperative games can be used to implement demand-side management or for deployment and control of microgrids~\cite{saad:12}.

The applicable game theoretic methods range from classical non-cooperative Nash games to advanced dynamic games.
For an application of game-theoretic methods in energy systems the following aspects have to be considered: First, the assumption that all players are perfectly rational might not hold. Decisions of real players in the smart grid scenario might deviate from the most rational solution due to failure, delay in learning, influence by other players, or a global information source. For example, if all neighbors are getting a photovoltaic power source, a house owner might be inclined to do so as well, with probable less regard of the pay-off expectation. A good advertising campaign for a certain product might also influence the player's decisions in a non-rational way.
Second, people might intentionally try to break the system -- either in order to increase their own revenue or just to cheat. Consider a smart microgrid network where a particular connection forms a distribution bottleneck during a short period of the day. Normally, market forces will overcome this bottleneck as good as possible, however, a \dq{cheating} player might deliberately block the access in order to create a supply shortfall with prices favorable for the player. Or, market players would deliberately place unrealistic bids in order to generate oscillations in the system.
Therefore, game-theoretic models and algorithms for smart microgrids are required to be robust against perturbations such as the impact of (deliberately or accidently) non-rational decisions. Fortunately, there exist methods such as the trembling hand equilibrium or strong time consistency that address these problems~\cite{young:04,basar:95,fudenberg:98}.

A use case for game theory is auction theory~\cite{avery:04} in order to negotiate on the price and amount of energy to be exchanged between networked microgrids. An introduction to possible auction models for smart micro-grids has been given earlier in this paper.
Another application of game theory would be to view the possibly cooperating smart micro-grids as players in a cooperative game. Several micro-grids can decide to form a coalition network that locally exchanges energy. Each micro-grid would try to optimize its payoff, which is a function of energy cost, energy loss due to transmission or over-production and fulfilment of energy needs of their customers. Apart from the static initial cost for networking the micro-grids, there would be a dynamic since with each new formation or modification of a coalition, the individual payoffs would change.
\section{Consumer Integration and Home Automation}
One of the hardest challenges of power systems is to balance demand and supply of energy. The advent of new energy consuming appliances such as electric vehicles make this task even harder. The goal is therefore to make consumers responsive by applying so called Demand Response (DR) mechanisms. Consumers should change their behavior in response to the market, by reducing usage during peak periods, by shifting their demand to off-peak periods or by using own generation facilities (e.g., photovoltaic systems). Different strategies can be implemented to encourage clients to actively participate, either by dynamic pricing, automatic load control or by incentive payments. In either way, the consumers need to be able to use their appliances without restrictions \cite{Albadi2008}. The interaction between consumer and grid will be enabled by smart meters, sensors, digital controls and analytical tools \cite{Jiang2011}. The decisions are often supported by automatic predictions or automatic schedules for home appliances and energy resources (home automation). In comparison to the before mentioned price models the following works consider consumer response in a feedback loop.

Demand Response can be used to forecast future demands of consumers, e.g., by tracking the consumption, weather, etc. over a long time period and then learning from these patterns for similar days \cite{Luh2010}. 
The authors consider for their future work also to integrate real-time client feedback, however, argue that the learning mechanism can become very complicated.

Most works combine home automation and demand response to increase the benefit for the consumers. In \cite{Morganti2009} the authors define that appliances are competing agents for hot water and energy. The goal is to avoid energy peaks (that could lead to outages) by automatically managing the appliances. One of the important points is that the consumers's actions are prioritized, e.g., if the consumer takes a shower and needs hot water, the dishwasher (if not heating water automatically) has to compete with the shower. During this time the dishwasher might be delayed, but the shower cannot. Based on these assumptions the authors use genetic algorithms. A set of parameters are considered as members of a chromosome. A number of chromosomes are regarded as population and by evaluating the chromosomes according to their predefined fitness, the next population is generated via reproduction, crossover and mutation. In comparison to traditional optimization the genetic algorithm is a heuristic approach that needs a number of generations to lead to good results. The authors claim that the design of the genetic algorithm has a high impact on its performance. The work shows promising results, however, the scope of action for a client had been severely restricted to reduce the search space.

In \cite{Negenborn2008} home-automation is made explicitly interactive. The proposed system recommends how a consumer can instantly save energy or costs. The authors call the system Model Predictive Control (MPC). MPC considers different control steps executed periodically (e.g., every 15 minutes). At these control steps the current system state is measured and based on that the optimization problem solved. The outcome is a number of possible actions to improve the system's performance, so called \dq{predictions} (of the energy savings).

A similar system is described by Pedrasa et al. in \cite{Pedrasa2010}. The authors automatically manage the home appliances by using particle swarm optimization (see section \ref{sec:pricing}). In this case particle swarms are operation schedules for the appliances, including the charging of an electric vehicle. For each appliance a number of swarms exist. However, best results are achieved if the schedules of the appliances are coordinated instead of considering them isolated. In this approach not only the energy consumption is optimized, but also the costs are reduced. The system targets to use battery stored power during high-price periods and to avoid photovoltaic energy export.

Client feedback should be as simple as possible, thus, there are solutions that let the client configure energy consumption tasks to occur during specific time periods. E.g., the dish washer should run between 9 a.m. and 5 p.m. Lightning and fridge have fixed consumption times that should not change. The task optimization is part of the home automation mechanism, such as introduced by \cite{Zhang2011}, \cite{Jiang2011}. Such a mechanism ensures that clients keep the control of their appliances and therefore the acceptance increases. Nonetheless, the prices might not be optimal in comparison to the full automatic scheduling.

In the following table \ref{tab:automation} we subsume the described approaches. We categorize the approaches by their (1) goal, i.e., prediction or feedback, (2) the optimization mechanism used and the (3) integration of the consumers. Fully automated solutions are considered passive, interactive solutions are considered as active user feedback.
\begin{table}
\centering
\caption{Comparison of home automation and consumer interaction, Optimization mechanims: prediction (pred.), feedback (fb.), mixed integer linear programming (milp), User Feedback: active (a), passive (p)}
\label{tab:automation}
\begin{tabular}{l l l l}
\\[-5pt]
\hline
  & Goal & Optimization Mechanism & User Feedback \\
   \hline
  \cite{Luh2010} & pred./fb. & - & p \\
  \cite{Morganti2009} & fb. & genetic algorithm & p \\
  \cite{Negenborn2008} & pred. & milp& a \\
  \cite{Pedrasa2010} &  pred./fb.& particle swarm optimization & p \\
  \cite{Zhang2011},\cite{Jiang2011} & pred./fb. & milp & a \\
  \hline
\end{tabular}
\end{table} 
\section{Load Balancing in Island Mode}
Operating in island mode is typically either motivated by high costs to access the grid or by the absence of a connection to the grid. In the latter case the disconnection is caused either by failure or to enable maintenance of the main grid without supply interruption. Some projects also employ intentional islanding to increase system reliability~\cite{katiraei:08, londero:10}.
Load balancing in microgrids comes with several challenges: While main distribution grids are typically significantly over-dimensioned regarding both their transmission capacity and their energy production flexibility, this is not the case for most microgrids. Load changes, for example, can be relatively large with respect to total load. This small scale of an islanded microgrid sets physical implications to load changes~\cite{Vandoorn2011}. Another challenge is the limited flexibility in energy generation. Alternative energy sources allow the spatial distribution of the energy production, but their output depends on environmental conditions (solar, wind). This makes it difficult to apply a load-following strategy, since the reserves on energy production are limit. Therefore, a smart microgrid will have to employ a load control strategy, which includes load shedding. In order to adjust to the time-of-day dependent energy generation performance, a demand dispatch strategy is motivated. To apply this kind of load control, a significant amount of the appliances must be \dq{smart}~\cite{elmenreich:wises12} in the way that they support load shedding or dispatching of their operating time.

Thus, we can identify the following methods for load balancing.
\begin{description}
\item[Generator power control:]
Assuming that the power generation of a smart microgrid cannot be controlled smoothly and fast enough in order to stabilize the microgrid voltage, droop control strategies are employed in order to keep the voltage within an acceptable band without changing the generation power\cite{Vandoorn2011}. Droop is the intentional loss in output voltage from a device as it drives a load by employing a series resistor between the regulator output and the load. Thus, the output voltage is less affected by load changes. Whenever the microgrid voltage exceeds its defined tolerance band, the generated power is adapted.
\item[Demand dispatch:]
Demand dispatch is a cooperative method to shift the energy consumption to a later point in time. Therefore, cooperation between smart devices and the microgrid power control is necessary.
Demand dispatch has been implemented before in the form of different energy tariffs for different times of the day, whereof the nighttime tariff was the most popular one. With the limited ability to handle peak load in microgrids, demand dispatch becomes more and more important.
Candidates for demand dispatch are dishwashers, washers and dryers, electric hot water heaters, HVAC systems with thermal storage, defrost cycles of refrigerators, battery, plug-in vehicles~\cite{lu:10}. Lu et al. estimated that \emph{33\% of all loads could have at least some level of demand dispatch control without a significant impact on end users}~\cite{lu:10}. For a microgrid the time scale of the load control can be very demanding, ranging from hours to even sub-seconds.
\item[Load shed:]
In case the load is exceeding the maximum generation power of the microgrid (or, the generation of additional power would not be economically feasible), a smart microgrid has to shed some load in order to keep the stability of the microgrid voltage. Load shed can be done by turning off devices or by reducing their power consumption, e.g., by operating them at a lower level of service. An example could be the heating of water, which, if taken back only a few degrees can safe significant amounts of energy at the cost of user's convenience. For turning off devices, a prioritization of devices has to be done based on the potentially reduced energy and their contribution to the overall level of service for the user. The problem of selecting the optimal set of devices for reducing a given amount of power consumption is equivalent to the Knapsack problem, which, when needed to be solved exactly, is NP-complete~\cite{martello:90}.
\item[Storage:]
With a storage unit, energy can be stored during times when production exceeds consumption. At times when consumption exceeds production, the storage can provide the missing demand. Thus, without the cooperation of devices or restricting the user the actual demand on energy generation can be shifted.
However, with current technologies, storage mechanisms for microgrids are expensive, especially when based on battery technology. Other storage approaches like pumped-storage hydroelectricity provide higher capacity and lower cost per energy unit, but do not scale down to the needs of a typical microgrid. Plug-in hybrid or electric cars might provide feasible energy storage for balancing smart grids without extra cost.
\end{description}

\begin{figure}[b]
  \includegraphics[width=\textwidth]{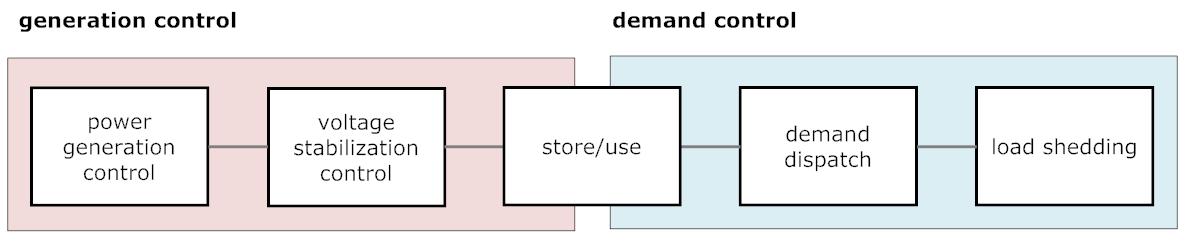}\\
  \caption{Load balancing methods}\label{fig:load_balancing_methods}
\end{figure}
These approaches are not exclusive but can be integrated for microgrid stabilization in island mode as depicted in Figure~\ref{fig:load_balancing_methods}. The different controllers forming a chain of supply and demand need to be tuned well in order to avoid instabilities such as a bullwhip effect in the power control.

 \section{Conclusion and Outlook}
\begin{figure}[b]
 \centering
  \includegraphics[scale=0.4]{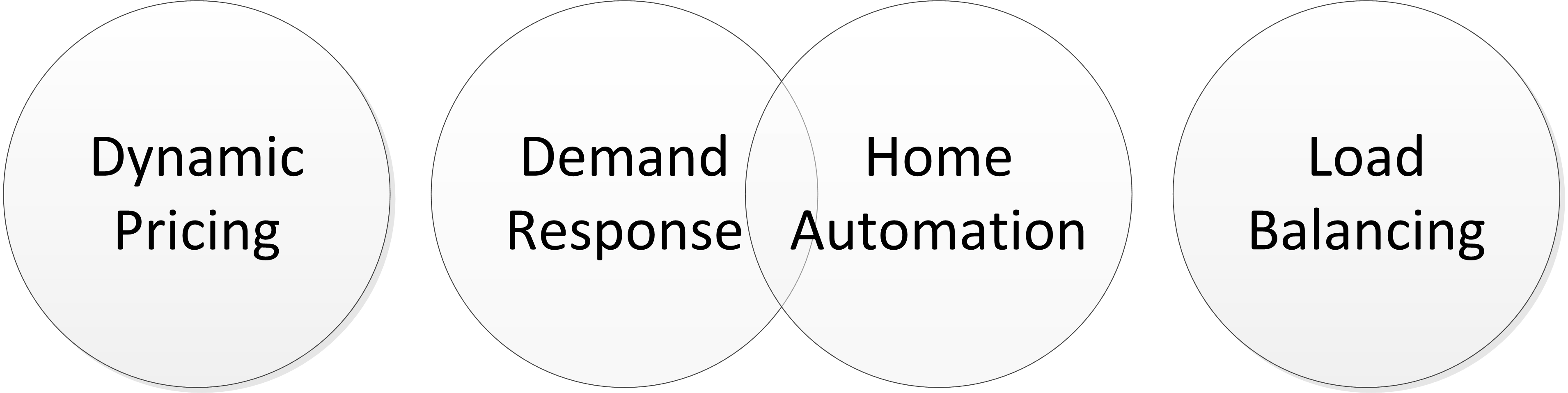}\\
  \caption{Current Research Efforts towards Smart Microgrids}
  \label{fig:current}
\end{figure}
We gave an overview on the main topics as shown in Figure \ref{fig:current}. The research efforts we subsumed are, except demand response and home automation, isolated. In demand response the authors often consider pricing as incentive for consumers to actively response to their energy consumption, however, the price and auction models already implemented by other researchers are not considered. The same applies for load balancing.
Initiatives such as the E-Energy projects in Germany\footnote{http://www.e-energy.de/en/} and the smart cities projects in Austria\footnote{http://www.smartcities.at/} show the necessity of integrating these efforts to holistic architectures.

\begin{figure}[t]
 \centering
  \includegraphics[scale=0.4]{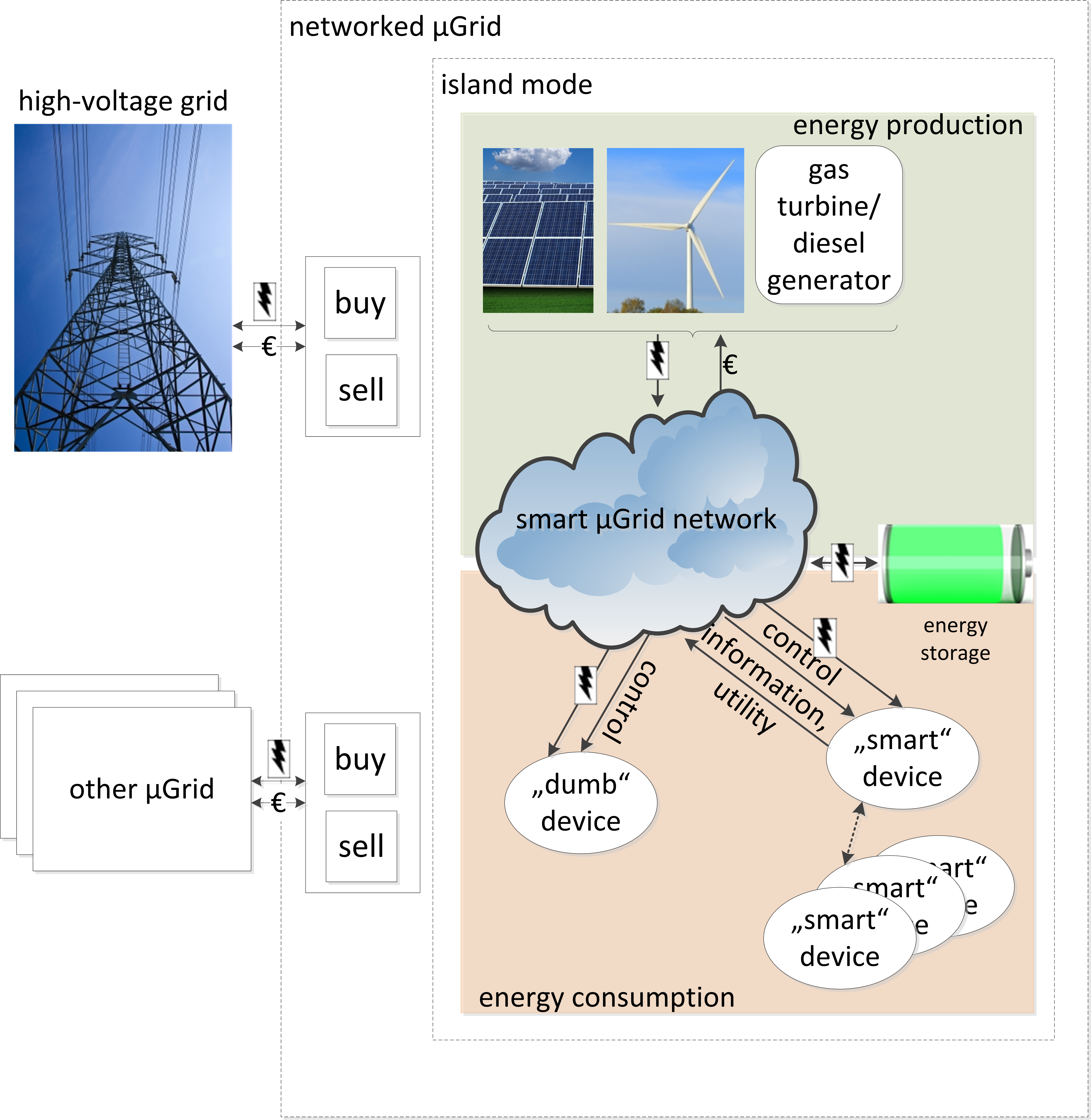}\\
  \caption{Outlook on networked smart microgrids }
  \label{fig:future}
\end{figure}

In the subsumed research topics of the above sections we have seen that many parts of a smart microgrid can be modeled as an optimization problem. Whenever the problem gets too complex, traditional methods or evolutionary algorithms are applied. On this basis we discuss possibilities of integrating these approaches to a whole architecture (see Figure \ref{fig:future}).

We have seen that for dynamic pricing the continuous double auction (CDA) principle is most common, but with different adaptations. The adaptations consider the optimization of either the market or a combination of load and market stability. To integrate the current load of the system an interface is needed to the consumers (DR-layer), because they might react due to the changing prices. Additionally, the energy load mechanisms need to be considered. The optimization on the level of the dynamic pricing layer of the architecture might get too complex and thus might need too long to be efficiently applied in a real system. Furthermore, full automated systems might not be accepted by the consumers, because they have no impact on how the system reacts.
In Figure \ref{fig:future}\footnote{Image references: David Castillo Dominici, Pixomar, dan, Sura Nualpradid / FreeDigitalPhotos.net} we pinpoint interactions between different actors of a smart microgrid. In islanded mode the microgrid has to rely on renewable energy sources, an emergency power generator and a storage facility (e.g., a battery). It consists of a number of devices, that are either "smart" or "dumb". Smart devices can be scheduled because of user interactions or implicit demands. It has to be considered that smart devices can be dependent on each other. The incentive mechanism is the price-based energy market. One can see that the interface to the high-voltage grid also interfaces the market. A microgrid consists of a buyers and sellers, whereas each of the energy sources can be a seller. The power generator collects information from devices and consumer feedback and performs auction based methods such as CDA. Since we assume that investments into power generation have been already done, we use the marginal costs to define the feasibility of using a particular generation unit. In the case of renewable energy sources, the available energy often comes with practical no marginal cost (indicated by \dq{\$ = 0} in the figure). In contrast, a diesel generator defines its marginal cost by the amount of fuel consumed during operation. For the consumer side, it is necessary to define a level of service and a relation of the level of service to some cost model. The optimization mechanism should maximize the consumer's level of service and use information on load, price, congestion, current renewable energy sources and the battery. Maximizing level of service could mean that a consumer can consume energy instantly when needed and the rest of the appliances are scheduled according to (1) network load and (2) energy price. Given that some appliances only provide a level of service in combination with other appliances (for example powering a computer monitor only makes sense if the computer is also running), this optimization is typically a complex task. Since decisions are created at different time scales and levels of criticality, there is a need for a structured and efficient energy optimization. Another open question, which we are planning to target in future work, is the communication between microgrids, e.g., for exchanging distributed energy resources.
\section*{Acknowledgments}
\small This work was supported by Lakeside Labs GmbH, Klagenfurt, Austria and funding from the European Regional Development Fund and the Carinthian Economic Promotion Fund (KWF) under grant KWF-$20214|22935|24445$.
\normalsize
 \bibliography{itg}
 \bibliographystyle{styles/lni}
\end{document}